\documentstyle[12pt,fleqn]{article}
\def\ra{\rightarrow}
\def\be{\begin{equation}}
\def\ee{\end{equation}}
\def\bea{\begin{eqnarray}}
\def\eea{\end{eqnarray}}

\begin{document}
\titlepage
\begin{flushright}IFT-2003-27\end{flushright}

~
\vspace*{0.5cm}

\begin{center}
{\Large\bf Photoproduction of isolated photon and jet at the DESY HERA}
\vspace*{3cm}

{\large Andrzej Zembrzuski, Maria Krawczyk}\\~

{\sl Institute of Theoretical Physics, Warsaw University,\\ 
ul. Ho\.za 69, 00-681 Warsaw, Poland}

\end{center}
\vspace*{3cm}

\begin{abstract}
The next-to-leading order (NLO) QCD calculation for the isolated photon 
and isolated photon plus jet photoproduction at the $ep$ collider 
DESY HERA is presented. The predictions for the isolated photon with
no restrictions imposed on the jet are compared with the previous
ones obtained in the small cone approximation, and the differences
are found to be below 2\%. The theoretical uncertainties in
the cross section of the photoproduction of the photon 
plus jet are discussed.
A short comparison with the new preliminary H1 data and with
the NLO predictions of Fontannaz et al. is also presented.
\end{abstract}
\thispagestyle{empty}
\newpage

\section{Introduction}

In the previous papers~\cite{Krawczyk:1998it,Krawczyk:2001tz} we 
have presented the next-to-leading order QCD calculation (NLO) for the 
photoproduction of a photon with large transverse momentum, 
so called prompt photon, at the $ep$ collider HERA. 
We have also shown the leading order (LO) predictions for the 
prompt photon and a jet production~\cite{Krawczyk:2001tz}~\footnote{A 
rough estimation of NLO predictions for the 
photon plus jet production was given in~\cite{Krawczyk:1998it}.}.

For comparison with experimental data, the final photon 
was considered in~\cite{Krawczyk:1998it,Krawczyk:2001tz} 
as an isolated one, i.e. the hadronic energy within some cone 
around the photon was restricted. We introduced the isolation 
in an approximated way with the assumption that the momenta of 
partons inside the isolation cone were almost collinear with the photon 
momentum (small cone approximation). 

\begin{figure}[b]
\label{comptonjet}
\vskip 5.2cm\relax\noindent\hskip 0cm
       \relax{\includegraphics{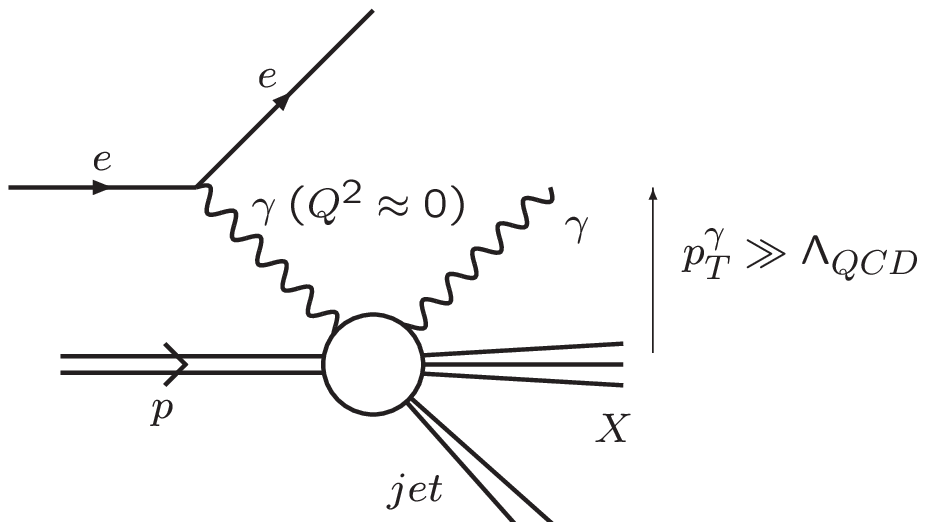}}
\vspace{0cm}
\caption{\small\sl The $ep\ra e\gamma ~jet ~X$ photoproduction.}
\label{fig:nojet.ptgamma}
\end{figure}

In this paper we study in NLO the photoproduction of both 
the isolated (prompt) photon and isolated (prompt) photon plus
a jet without the small cone approximation. Since the numerical 
differences between the present calculation and the previous 
approximated one are found to be very small for the photoproduction of
the isolated photon, we focus our attention mainly 
on the NLO predictions for the isolated photon observed in the 
final state together with a jet, see fig.~\ref{comptonjet}.

The photoproduction is a process in which a real or almost real
photon is collided with another particle in order to produce
some final state
(for a recent reviews see~\cite{Klasen:2002xb,Krawczyk:2000mf,Nisius:2000cv}).
In an electron-proton scattering the photoproduction corresponds
to the small virtuality of the exchanged photon, $Q^2$, say $Q^2$
\begin{minipage}[t]{10pt} \raisebox{3pt}{$<$} \makebox[-17pt]{}
\raisebox{-3pt}{$\sim$}\end{minipage} 1 GeV$^2$
\cite{Breitweg:1997pa,unknown:uj,Breitweg:1999su,Chekanov:2001aq,h12003}. 
In such a case this almost real photon colliding with the proton
may lead for instance 
to a production of jets or particles with large transverse momenta.

The photoproduction of the prompt photon in the $ep$ or $\gamma p$ collision
have been studied theoretically (for a non-isolated final photon) since 
1969~\cite{Bjorken:1969ja}-\cite{Gordon:1994sm}.
Then the NLO predictions for the isolated photon have been
presented by various authors~\cite{Gordon:1995km,Krawczyk:1998it,
Fontannaz:2001ek,Krawczyk:2001tz}.
The photoproduction of the isolated photon and a jet at HERA have been
calculated in NLO by Gordon~\cite{Gordon:1998yt}
and Fontannaz, Guillet and Heinrich~\cite{Fontannaz:2001nq}.
There is also a NLO calculation for the isolated photon
and isolated photon plus a jet production in the deep inelastic
scattering ($Q^2\gg 1$ GeV$^2$)~\cite{Kramer:1998nb},
which is in many theoretical aspects close to the calculations
for the corresponding photoproduction events.

The photoproduction of isolated photons without and with additional jets 
has been measured at the HERA collider by the ZEUS~\cite{Breitweg:1997pa,
unknown:uj,Breitweg:1999su,Chekanov:2001aq} and H1~\cite{h11997,h12003}
Collaborations. 
In general, the data are in reasonably agreement with NLO predictions.
Nevertheless, none of theoretical predictions as well as
Monte Carlo simulations give sufficiently good description of the ZEUS 
data~\cite{unknown:uj,Breitweg:1999su} (see also~\cite{Bussey2001})
for the isolated photon at negative photon rapidities ($\eta_{\gamma}$).
Some discrepancies between NLO~\cite{Fontannaz:2001nq} and data are 
also seen for the preliminary H1 data~\cite{h12003}, mainly at 
$\eta_{\gamma} > 0.6$ and $x_{\gamma}<0.5$.

In the paper~\cite{Chekanov:2001aq} the ZEUS Collaboration has 
considered in Monte Carlo simulations the intrinsic transverse 
momentum of partons in the proton in photoproduction events 
containing the prompt photon plus
jet, and the effective intrinsic transverse momentum
was found to be $<k_T>=1.69\pm0.18^{+0.18}_{-0.20}$ GeV. 
On the other hand, no need for the intrinsic transverse momentum 
in the proton was reported in~\cite{Fontannaz:2001nq}.

In our analysis the intrinsic transverse momentum is
not included, since we think that the measured $<k_T>$ 
is not due to the transverse momenta of partons inside the proton, but it 
describes effectively higher order emissions or multiple interactions.

The paper is organized as follows.
First, in sec~\ref{nlo} the processes contributing in NLO and
the general formula for the cross section are shortly presented.
The isolation cuts and the jet definition 
are introduced in sec.~\ref{isol}.
In sec.~\ref{pss} the slicing of the three body phase space
is discussed, and some other calculation details and inputs are given
in sec.~\ref{det}.
The predictions for the $ep\ra e\gamma X$ process are given
in sec.~\ref{nojet}. The results for the
$ep\ra e\gamma ~jet ~X$ process, discussion of theoretical
uncertainties, and comparison with the new preliminary
H1 data, as well as with the predictions 
of Fontannaz et al.~\cite{Fontannaz:2001nq} are presented in sec.~\ref{jet}.
A short summary is given in sec.~\ref{sum}.

\section{Cross section}\label{nlo}

In our previous papers~\cite{Krawczyk:1998it,Krawczyk:2001tz} 
we have discussed the NLO calculation of the cross section of the 
isolated photon production in the $ep$ collision.
The set of contributing diagrams was different than in NLO calculations
of other authors~\cite{Gordon:1995km,Fontannaz:2001ek,
Gordon:1998yt,Fontannaz:2001nq}, since we assumed that parton densities
in the photon were of order $\mathcal{O}$$(\alpha )$, 
while in the cited papers their were assumed to be of order 
$\mathcal{O}$$(\alpha /\alpha_S)$.

The parton densities inside the photon are proportional to the
electromagnetic coupling, $\alpha$, and to the logarithm of the energy 
scale $\mu$ at which the densities are probed, 
$\ln (\mu^2/\Lambda_{QCD}^2)$~\footnote{We define the scale 
$\mu$ in sec.~\ref{det}.}~\cite{Walsh:1973mz,DeWitt:1978wn,Chyla:1999mw}.
This logarithm is formally proportional to $1/\alpha_S (\mu^2)$ but
it originates from the pure electromagnetic process,
$\gamma\ra q\bar{q}$, 
with no strong interactions, and therefore the parton densities
in the photon should be treated as quantities of order 
$\mathcal{O}$$(\alpha )$, 
and not $\mathcal{O}$$(\alpha /\alpha_S)$.
This different treatment
of parton densities leads to different set of diagrams
in NLO calculations involving resolved photons, and to some 
moderate differences 
in numerical predictions in comparison with predictions of other 
authors~\cite{Krawczyk:2001tz}.

In the present calculation we take into account the same set of 
contributing partonic processes as it was taken in~\cite{Krawczyk:2001tz},
namely: the Born diagram ($\gamma q\ra\gamma q$) together with QCD 
corrections, the box diagram ($\gamma g\ra\gamma g$)~\cite{Combridge:1980sx},
processes with the resolved initial or final photon, and processes
with resolved both the initial and final photon.
The general formula for the differential cross 
section for the $ep\ra e\gamma X$ (or $ep\ra e\gamma ~jet ~X$) 
photoproduction is:
\be\label{epa}
d\sigma^{ep\ra e\gamma (jet) X} = \int G_{\gamma /e}(y)
d\sigma^{\gamma p\ra \gamma (jet) X} dy ~,
\ee
and
\bea
d\sigma^{\gamma p\ra \gamma (jet) X}=
\sum_{a=\gamma ,q,\bar{q},g} \int dx_{\gamma} 
\sum_{b=q,\bar{q},g} \int dx 
\sum_{c=\gamma ,q,\bar{q},g} \int {dz\over z^2}
f_{a/\gamma}(x_{\gamma},\mu^2)\cdot
\nonumber
\eea
\be\label{cross}
\cdot 
f_{b/p}(x,\mu^2)
D_{\gamma /c}(z,\mu^2)
d\sigma^{ab\ra cd_1} +
\sum_{b=q,\bar{q},g} 
\int dx f_{b/p}(x,\mu^2) d\sigma^{\gamma b\ra\gamma d_1d_2} ~,
\ee
where $ab\ra cd_1$ and $\gamma b\ra\gamma d_1d_2$ are the partonic processes.

The eq.~\ref{epa} is the Weizs\"{a}cker-Williams
approximation~\cite{vonWeizsacker:1934sx,Williams:1934ad}
(see also e.g.~\cite{Budnev:1974de,Frixione:1993yw,Nisius:2000cv})
with the flux of the real photons emitted from the electron given
by $G_{\gamma /e}(y)$, where $y$ is the fraction of the initial electron
momentum carried by the photon. The functions $f_{a/\gamma}$, 
$f_{b/p}$, and $D_{\gamma /c}$ stand for parton distributions
in the photon and proton, and a fragmentation function into the photon, 
respectively.
For the direct initial ($a=\gamma$) or final ($c=\gamma$) photon
the corresponding distributions are replaced by the $\delta$-functions:
$f_{a/\gamma}=\delta(x_{\gamma}-1)$ or $D_{\gamma /c}=\delta(z-1)$.
The longitudinal-momentum fractions in the photon and proton parton
densities, and in the
fragmentation functions are denoted as $x_{\gamma}$, $x$, and $z$,
respectively, while $\mu$ stands for the factorization/renormalization scale.

The calculation of the partonic cross sections,
$d\sigma^{\gamma b\ra\gamma d_1(d_2)}$, involve isolation
restrictions and other kinematic cuts, as used in
corresponding experimental analyses.

\section{Isolation cut and jet definition}\label{isol}

The NLO calculation includes partonic processes with two ($2\ra 2$) or three
($2\ra 3$) particles in the final state:
\be\label{223}
ab\ra cd_1(d_2)\nonumber ,
\ee
where $a$ is the photon or a parton originating from the photon, 
$b$ is a parton from the proton, $c$ stand for the final photon 
or a parton which decay into the photon in the fragmentation process, 
and $d_i$ are quarks and/or gluons. 
The outgoing partons $d_i$ are not observed in the final state 
-~they recombine into colorless hadronic jets.

In experimental analyses the final photon is required to be isolated,
i.e the transverse energy of hadrons inside a cone of radius R around 
the photon is assumed to be less than $\epsilon$ times the photon 
transverse energy,
\be\label{eps}
\sum_{hadrons} E_T^{hadron} < \epsilon E_T^{\gamma} ,
\ee
where $\epsilon$ is a small parameter.
The cone is defined in the rapidity and azimuthal-angle phase space,
\be\label{R}
\sqrt{(\eta_{hadron}-\eta_{\gamma})^2+(\phi_{hadron}-\phi_{\gamma})^2} < R.
\ee

The some isolation restriction is taken into
account in theoretical parton-level predictions, where the 
sum in eq.~(\ref{eps}) runs over $c$-parton remnant and
over $d_i$-partons, if they are inside the cone (\ref{R}). 
In the present calculation we take $R=1$ and $\epsilon =0.1$,
as in the H1 and ZEUS Collaborations' analyses.

In the partonic $2\ra 2$ processes with a direct final photon ($c=\gamma$),
\be\label{22nofrag}
ab\ra\gamma d_1 ,
\ee 
the photon is isolated by definition.
If the final photon comes from the fragmentation process ($c\ne\gamma$),
\be
ab\ra c d_1 ,
\ee 
it is isolated for the momentum fraction $z>1/(1+\epsilon)$,
and this cut is used in an integration in eq.~(\ref{cross}).
An inclusion of the isolation in the partonic cross sections
for $2\ra 3$ processes, 
\be
\gamma b\ra\gamma d_1 d_2 ,
\ee
is more complicated, since one needs to restrict the final momenta 
of two partons (eq.~\ref{eps},~\ref{R}), 
and to care about a cancellation 
and factorization of singularities, as described in the next section.

For $2\ra 2$ processes only one jet appears,
and it originates from the $d_1$-parton.
The jet's transverse energy, rapidity, and azimuthal angle
are assumed equal to the transverse energy,
rapidity, and azimuthal angle of the $d_1$-parton.

Two partons, $d_1$ and $d_2$, produced in the $2\ra 3$ subprocess 
may lead to two separate
jets or they may form one jet. A number of jets in the final state depends
on the jet definition. In this paper the inclusive $k_T$-jet finding 
algorithm~\cite{Ellis:tq} is employed.

Since in NLO calculation we deal with no more than two partons
forming a jet/jets,
the algorithm becomes very simple. If the distance between the partons,
$R_{12}$, defined as
\be
R_{12}=\sqrt{(\eta^{d_1}-\eta^{d_2})^2+(\phi^{d_1}-\phi^{d_2})^2} ~,
\ee
is larger than an arbitrary parameter $R_J$,
then two separate jets arise with transverse energies, rapidities, 
and azimuthal angles of the $d_i$-partons:
\be
E_T^{jet_i}=E_T^{d_i}\makebox[1cm]{,}
\eta^{jet_i}=\eta^{d_i}\makebox[1cm]{,}
\phi^{jet_i}=\phi^{d_i}\makebox[1cm]{,}
i=1, ~2.\nonumber
\ee
For $R_{12} < R_J$ the $d_i$-partons are treated as ingredients of
one jet with
\be
E_T^{jet}=E_T^{d_1}+E_T^{d_2} ,\label{jaE}
\ee
\be
\eta^{jet}=(E_T^{d_1}\eta^{d_1}+E_T^{d_2}\eta^{d_2})/E_T^{jet}\label{jan} ,
\ee
\be
\phi^{jet}=(E_T^{d_1}\phi^{d_1}+E_T^{d_2}\phi^{d_2})/E_T^{jet}\label{jap} .
\ee
Following experimental analyses,
$R_J=1$ will be used in numerical calculations.

\section{Phase-space integration}\label{pss}

To obtain predictions, one need to perform an integration over 
four-momentum of at least one final particle, since 
unintegrated partonic cross sections contain $\delta^{(4)}$-functions.
In case of $2\ra 2$ processes with no virtual gluon exchange
this integration is straightforward, and will not be discussed.

There are three types of diagrams giving NLO corrections to the Born process, 
namely:
$\gamma q\ra\gamma qg$ (real gluon emission), $\gamma g\ra\gamma q\bar{q}$,
and the process $\gamma q{\stackrel {g^*}{\longrightarrow}}\gamma q$ 
with a virtual gluon exchange.
The cross sections of the above processes contains infrared
singularities. For numerical integration over momenta 
of final particles, one needs to isolate these 
singularities~\cite{Fontannaz:2001ek,Gordon:1998yt}.
Like in earlier calculations for the $ep\ra e\gamma (jet) X$ 
reaction~\cite{Fontannaz:2001ek,Gordon:1998yt,Fontannaz:2001nq},
also in the present one the phase space is sliced into a few parts.

Let us assume that 
$\theta_{ji}$ is an angle between the momentum of $d_i$ parton
and the momentum of an initial particle $j$, where $j=e,~p$.
Let us also define the distance, $R_{\gamma i}$, between the 
$d_i$ parton and the final photon:
\be
R_{\gamma i}=\sqrt{(\eta_{d_i}-\eta_{\gamma})^2+(\phi_{d_i}-\phi_{\gamma})^2}.
\ee

We define the parts of the phase space in the following way:

$\bullet$ Part 1. In the first part the variable $w$ 
(see Appendix~\ref{notation})
is assumed in the range $w_{cut} \le w \le 1$, where $w_{cut}$ 
is an arbitrary parameter close to 1: $0<1-w_{cut}\ll 1$. 
This region of the phase space contains
all types of NLO corrections: virtual gluon exchange, real gluon
emission, and the process $\gamma g\ra\gamma q\bar{q}$.
The virtual gluon exchange is a $2\ra 2$ process, while the other
ones are of $2\ra 3$. However, for $w$ close to 1, $w_{cut} < w \le 1$,
the two final partons in the $2\ra 3$ processes are almost collinear
or/and one of the final partons is soft.
For $w_{cut}$ sufficiently close to 1 the kinematics of
the $2\ra 3$ processes is almost the same as in the $2\ra 2$ case:
\be
\gamma b\ra\gamma d \makebox[1cm]{,}d\equiv d_1 + d_2,
\ee
where the ``particle'' $d$ has four-momentum equal to the sum
of four-momenta of the $d_i$-partons, and its mass is almost zero.
In this case the final photon is isolated
and the jet can be identified with the $d$-``particle''.

The other parts of the phase space, parts 2-5 described below, 
are defined for $w<w_{cut}$, and contain only $2\ra 3$ processes.

$\bullet$ Part 2. In this region $w<w_{cut}$ and 
$\theta_{cut} > min (\theta_{e 1}, \theta_{e 2})$,
where $\theta_{cut}$ is a small arbitrary parameter, $\theta_{cut}\ll 1$.
Here, one of the final partons has the 
momentum almost collinear to the momentum
of the initial electron and it, for sufficiently small $\theta_{cut}$,  
does not enter the isolation cone around the final photon,
and does not enter the cone defining the jet.
The second parton has a large transverse momentum balancing the
photon transverse momentum, so the final photon 
is isolated, and the jet consists (on the 
partonic level) of only this very parton.

$\bullet$ Part 3. Here $w<w_{cut}$ and 
$\theta_{cut} > min (\theta_{p 1}, \theta_{p 2})$.
One of the final partons has momentum almost collinear to the momentum
of the proton, and, like in part 2, the final photon is isolated
and the high-$E_T$ jet originates from the second parton alone.

$\bullet$ Part 4. Here $w<w_{cut}$ and 
$R_{cut} > min (R_{\gamma 1}, R_{\gamma 2})$,
where $R_{cut}$ is a small parameter, $R_{cut}\ll1$.
In this case one of the final partons, say $d_1$-parton, is almost collinear 
to the final photon and the photon is isolated if 
$E_T^{d_1}<\epsilon E_T^{\gamma}$. This $d_1$-parton does not contribute to
the jet.

$\bullet$ Part 5. The last part is defined as the region in which there
are no collinear configurations: $w<w_{cut}$, 
$\theta_{cut} < min (\theta_{e 1}, \theta_{e 2})$,
$\theta_{cut} < min (\theta_{p 1}, \theta_{p 2})$, and
$R_{cut} < min (R_{\gamma 1}, R_{\gamma 2})$.

In numerical calculations we apply in part 1
the formulae integrated over four-momenta of
the final partons and virtual gluons~\cite{Aurenche:1984hc,jan},
namely the eqs.~(24) and (37) from ref.~\cite{Aurenche:1984hc}. 
In these formulae all the soft gluon singularities present in the virtual 
gluon and real gluon corrections are canceled, and all the
collinear singularities are factorized into the parton densities.

Parts 2-5 contain no contribution from the virtual gluon exchange,
and no soft gluon singularities. 
Analytical expressions for cross sections corresponding to
parts 2, 3, and 4 are given in the Appendix~\ref{col}. All the collinear
singularities appearing in the calculations are factorized
into parton densities in the photon (part 2), proton (part 3),
and into fragmentation functions (part 4).

The cross section in part 5 has no singularities at all, 
and one can perform an exact numerical integration over
final four-momenta in any kinematic range, including isolation 
restrictions and other cuts.
For this purpose we use the squared matrix element given in eqs.~(36) 
(with $\varepsilon =0$) in ref.~\cite{Aurenche:1984hc}.

We assume that the cut-off angle $\theta_{cut}$ is defined 
in the centre of mass of the initial photon and the initial parton.
The predictions should not depend on a choice of unphysical cut-off parameters,
$1-w_{cut}$, $\theta_{cut}$ and $R_{cut}$, if they are small enough. 
On the other hand, they can not be too small, since very low values lead
to large numerical errors.
The results presented in secs.~\ref{nojet} and ~\ref{jet} are obtained
with $1- w_{cut}=\theta_{cut}=R_{cut}=0.01$.
We have checked that the change of the predictions due to the variation
of these parameters is negligible, below 1\%, if they are taken
in the range $10^{-4} \le 1 - w_{cut} \le 0.03$, 
$10^{-4} \le \theta_{cut} \le 0.05$ and
$10^{-4} \le R_{cut} \le 0.05$.

\section{Other calculation details}\label{det}

Our NLO calculations are performed in the $\overline{\rm MS}$ scheme.
The factorization/renor\-malization scales in parton densities and 
fragmentation functions are assumed equal to the
renormalization scale in the strong coupling constant and are denoted
as $\mu$. As a reference $\mu$ equal to the transverse
momentum (transverse energy) of the final photon is taken,
$\mu=p_T^{\gamma}=E_T^{\gamma}$. For comparison $\mu=E_T^{\gamma}/2$ 
and $\mu=2E_T^{\gamma}$ will be also considered.

The two-loop strong coupling constant,
\be
\alpha_S(\mu^2)={{12 \pi}\over {(33-2N_f)\ln(\mu^2/\Lambda^2)}}
[1-{{6(153-19N_f)}\over (33-2N_f)^2} {{\ln[\ln(\mu^2/\Lambda^2)]}
\over{\ln(\mu^2/\Lambda^2)}}] ~,
\ee
is used with the QCD parameter $\Lambda$=0.386, 0.332 and 0.230 GeV 
for the number of active massless flavours $N_f$=3, 4 and 5,
respectively. In LO we use one-loop $\alpha_S$ with $\Lambda$=0.123 GeV
and $N_f$=4. The above $\Lambda$ values are obtained by us from 
the world average of $\alpha_S$ at the scale 
$M_Z$~\cite{Bethke:2002rv}:
\be
\label{1183}
\overline{\alpha_S} (M_Z) = 0.1183 \pm 0.0027.
\ee
To minimize theoretical and experimental uncertainties,
$\overline{\alpha_S} (M_Z)$ was determined from precise data based on 
NNLO analyses only. The data given at scales different than $M_Z$ were 
extrapolated to the $M_Z$ scale using the four-loop coupling. Although we
use one- and two-loop expressions, we have applied~(\ref{1183})
as the best estimation of the true value of $\alpha_S (M_Z)$.

In numerical calculations we take the Gl\"uck-Reya-Vogt (GRV)
LO and NLO parton densities in the proton~\cite{Gluck:1995uf}
and photon~\cite{Gluck:1992ee}, and the NLO GRV
fragmentation functions~\cite{Gluck:1993zx}. For comparison 
we also use other parametrizations, namely
Martin-Roberts-Stirling-Thorne (MRST2002)~\cite{Martin:2002aw},
CTEQ6~\cite{Pumplin:2002vw}, Aurenche-Guillet-Fontannaz
(AFG)~\cite{Aurenche:1994in} and (AFG02)~\cite{afg02}, 
Gordon-Storrow (GS)~\cite{Gordon:1997pm}, 
Cornet-Jankowski-Krawczyk-Lorca (CJKL)~\cite{Cornet:2002iy},
Duke-Owens (DO)~\cite{Duke:1982bj}, and
Bourhis-Fontannaz-Guillet (BFG)~\cite{Bourhis:1997yu}.

The initial electron and proton energies at the HERA collider
are assumed equal to
$E_e=27.6$ GeV and $E_p=920$ GeV, respectively.
In the Weizs\"{a}cker-Williams 
approximation~\cite{vonWeizsacker:1934sx,Williams:1934ad}
the photon spectrum in the electron is taken in the 
form~\cite{Frixione:1993yw}:
\be
G_{\gamma/e}(y)={\alpha\over 2\pi} \{ {1+(1-y)^2\over y}
\ln [ {Q^2_{max}(1-y)\over m_e^2 y^2}]
- ~ {2\over y}(1-y-{m_e^2y^2\over Q^2_{max}}) \} ~,
\ee
with the maximal virtuality $Q^2_{max} = 1$ GeV.
The above formula describes the spectrum of equivalent real
(transversally polarized) photons. We neglect longitudinally polarized
photons and the interference between longitudinally and transversally 
polarized photons, since they give very small 
contribution~\cite{Jezuita-Dabrowska:bp}.
As usual, we also do not take into account an emission of the final
large-$p_T$ photon directly from the electron~\cite{ula}.

The results presented in next sections are obtained in NLO QCD
with use of the GRV set of parametrizations 
with $\mu =E_T^{\gamma}$ and $N_f=4$ unless explicitly stated otherwise.

\section{Results for the $ep\ra e\gamma X$ process}\label{nojet}

The comparison with the ZEUS Collaboration 
data~\cite{Breitweg:1999su} for the photoproduction of isolated photons
have been discussed in details in~\cite{Krawczyk:2001tz}. Recently a new
preliminary data have been presented by the H1 Collaboration~\cite{h12003}.
For a comparison with these data, we apply 
kinematic cuts used in~\cite{h12003}, namely
the fraction of the electron energy transferred to the photon
is restricted to the range $0.2<y<0.7$, and the final photon rapidity
and transverse energy are taken in the limits
$-1<\eta_{\gamma}<0.9$ and $5<E_T^{\gamma}<10$~GeV, respectively.

In fig.~\ref{fig:nojet}
the differential cross sections $d\sigma /dE_T^{\gamma}$ and
$d\sigma /d\eta_{\gamma}$
are shown. Our exact predictions (solid lines) are compared
with the predictions obtained in the small cone approximation 
(dashed lines) discussed in our previous paper~\cite{Krawczyk:2001tz}.
The differences do not exceed 2\%, so the
small cone approximation works very well, despite the fact that the isolation
cone of radius $R=1$ is not a small one.

The predictions are in good agreement with the H1 preliminary 
data~\cite{h12003} (not shown) for both $d\sigma /dE_T^{\gamma}$ and
$d\sigma /d\eta_{\gamma}$ distributions with exception of one
experimental point at $\eta_{\gamma} > 0.58$, which lies slightly below 
the predictions.

\section{Results for the $ep\ra e\gamma ~jet ~X$ process}\label{jet}

There are two publications of the ZEUS Collaboration
presenting results of
measurements of the isolated photon plus jet photoproduction
at the HERA Collider~\cite{Breitweg:1997pa,Chekanov:2001aq}.
In the first paper~\cite{Breitweg:1997pa}
the total cross section integrated over some kinematic range 
is given. The aim of the second one~\cite{Chekanov:2001aq} was to study
transverse momentum of partons in the proton, and no data
for cross sections were presented (although some data
for distributions of events, not corrected for the detector
effects, were shown).

In the new paper of the H1 Collaboration~\cite{h12003} 
(see also~\cite{Lemrani:2003mj}), the
preliminary photoproduction data are presented for various 
differential cross sections of both $ep\ra e\gamma X$ (considered
above in sec.~\ref{nojet}) and $ep\ra e\gamma ~jet ~X$ processes.

As in sec.~\ref{nojet}, 
we impose kinematic limits used in the H1 analysis~\cite{h12003}.
The cross sections are integrated over $0.2<y<0.7$, and 
$-1<\eta_{\gamma}<0.9$ and/or $5<E_T^{\gamma}<10$~GeV
with the jet rapidity and jet transverse energy in the
range $-1<\eta_{jet}<2.3$ and 4.5 GeV $<E_T^{jet}$, respectively.
If two jets are found with the above parameters, that
with higher $E_T^{jet}$ is taken.

\subsection{Theoretical uncertainties}\label{res}

As it is discussed in details in~\cite{Fontannaz:2001nq}, the symmetric cuts
for the photon and the jet transverse energy leads to unphysical
results in next-to-leading or higher orders of
calculations. This effect is shown in fig.~\ref{fig:ptgamma}a, where
the LO and NLO predictions as a function of the photon
transverse energy are shown in the $E_T^{\gamma}$-range wider than 
the range considered by the H1 Collaboration.
At $E_T^{\gamma}$ values close to the minimal jet transverse energy,
$E_{T,min}^{jet} = 4.5$ GeV, the NLO differential cross section 
has a discontinuity: for $(E_T^{\gamma})_-\ra 4.5$ GeV it has 
a strong maximum, and a minimum for $(E_T^{\gamma})_+\ra 4.5$ GeV.
In the minimum the value of the cross section is even negative.
This unphysical fluctuation is not present in the LO calculation.

An integration of the differential cross section over the photon
transverse energy higher than the minimal jet transverse energy,
$E_T^{\gamma} \ge E_{T,min}^{jet}$, leads to underestimated 
predictions in NLO. However numerically this effect is not very important.

To avoid theoretical instabilities, one can consider a cross section
averaged over some $E_T^{\gamma}$-bins, see fig.~\ref{fig:ptgamma}b.
Note, that the NLO predictions are well defined if one takes
the cross section integrated/averaged over $E_T^{\gamma}$
from $E_{T,min}^{jet} - \Delta$ to $E_{T,min}^{jet} + \Delta$,
if $\Delta$ is sufficiently large, say $\Delta > 0.3$ GeV.
So, the bins of a length 1 GeV presented in fig.~\ref{fig:ptgamma}b
are large enough to avoid errors due to the symmetric cuts.

The cross section for $E_T^{\gamma}<E_{T,min}^{jet}$ is dominated
by the NLO corrections, since the LO contribution is suppressed by the
$E_T^{jet}$-cut (4.5 GeV $<E_T^{jet}$) 
and by the isolation requirement (sec.~\ref{isol}).
We checked that the dependence on the
re\-normal\-ization/factor\-ization scale, $\mu$, is not strong:
variations of $\mu$ from $E_T^{\gamma}$ to $E_T^{\gamma}/2$ or $2E_T^{\gamma}$ 
lead to changes of the cross section less than
3\% for $E_T^{\gamma}<E_{T,min}^{jet}$ and up to 5\% 
for $E_T^{\gamma}>E_{T,min}^{jet}$.

The NLO predictions for various $\mu$, and the LO predictions are also shown
in fig.~\ref{fig:jet}. In each presented here
bin the dependence on the choice of $\mu$ is less
than $\pm$5\% for both $E_T^{jet}$ and $\eta_{jet}$ distributions. 
The cross section is suppressed for $E_T^{jet}$
close to or higher than the maximal transverse energy of the photon,
$E_{T,max}^{\gamma}=10$ GeV. 
At negative $\eta_{jet}$ the LO predictions are higher by 14\%,
and at positive $\eta_{jet}$ they are lower than the NLO ones by 16-27\%.

Since for $E_T^{\gamma}>E_{T,min}^{jet}$
the difference between LO and NLO is not larger than 27\% and the 
variation of NLO results due to the variation
of the $\mu$ scale is small, up to 5\%, the calculation is stable, and
one can expect that the QCD corrections of higher orders
are not sizable. On the other hand, for $E_T^{\gamma}<E_{T,min}^{jet}$
the NLO corrections constitute almost 100\% of the
cross section, and the corrections of higher orders may change 
predictions.

In fig.~\ref{fig:etagamma} the results obtained using different 
parton densities in the proton are shown. The predictions of 
CTEQ6 (NLO)~\cite{Pumplin:2002vw} are 6\% higher than
the predictions of MRST2002 (NLO)~\cite{Martin:2002aw}. The 
GRV (NLO)~\cite{Gluck:1995uf} densities 
give results higher than MRST2002 by 5-7\% at negative $\eta_{\gamma}$,
and 3-5\% at positive $\eta_{\gamma}$. Differencies between CTEQ6
and GRV do not exceed 4\%.

The comparison between different parton densities in the photon
is shown in fig.~\ref{fig:x} for the $x_{\gamma}^{obs}$ distribution,
where $x_{\gamma}^{obs}$ is defined as
\be
x_{\gamma}^{obs}=(E_T^{jet}e^{-\eta^{jet}}+E_T^{\gamma}e^{-\eta^{\gamma}})
/2yE_e ~.
\ee
The GS (NLO) parametrization~\cite{Gordon:1997pm} give predictions
lower than GRV (NLO)~\cite{Gluck:1992ee} by 20-36\%
for $x_{\gamma}^{obs}<0.9$. This large difference is due to
the charm threshold assumed in the GS parametrization
at large scale, $\mu^2 = 50$ GeV$^2$, much higher than
thresholds in other considered herein 
parametrizations~\cite{Krawczyk:2001tz}.
The results obtained with use of AFG (NLO)~\cite{Aurenche:1994in} 
and AFG02 (NLO)~\cite{afg02} are very similar, and only 
the latter is shown in fig.~\ref{fig:x}. It gives predictions 
up to 15\% lower than GRV for $x_{\gamma}^{obs}<0.9$. At 
large-$x_{\gamma}^{obs}$ region, $x_{\gamma}^{obs}>0.9$, the 
cross section is dominated by the contribution of processes
with direct initial photons, and the differences between
predictions of various parametrizations are small.
For the total cross section integrated over all $x_{\gamma}^{obs}$
values the difference between GRV and AFG02 (GS) is 4\% (16\% ).

We have also compared predictions of GRV (LO)~\cite{Gluck:1992ee}
and the new CJKL (LO)~\cite{Cornet:2002iy} parton densities in the photon,
as well as predictions of DO (LO)~\cite{Duke:1982bj}, 
GRV (NLO)~\cite{Gluck:1993zx} and BFG (NLO)~\cite{Bourhis:1997yu} 
fragmentation functions. The total LO cross section (within
the kinematic range considered by the H1 Collaboration)
for the GRV (LO) parametrization is higher than for CJKL (LO) by 3\%.
The isolation requirement reduce the contribution of processes
involving the resolved final photon~\cite{Krawczyk:2001tz}, 
so the dependence on the choice of fragmentation functions is weak, even if
the fragmentation functions differ considerably from one another.
The total cross sections obtained with DO and BFG (set I and set II)
are lower than the predictions of GRV by 2\% and 4\%, respectively.

The GRV distributions for the proton, photon and fragmentation
have been used as a reference,
and each time only one parametrization has been changed
for a comparison. The differences observed in the total cross
section are not large (with an exception of the GS densities, which
give predictions considerably lower than the other densities
in the photon due to the specific treatment of the charm contribution). 
However, the differences can be larger if
one changes simultaneously all the used distributions.
For instance predictions of the MRST, AFG02 and BFG set
are lower than the GRV predictions by about 11\% .

\subsection{Comparing with H1 preliminary data
and with FGH predictions}\label{h1}

The predictions shown in figs.~\ref{fig:ptgamma}-\ref{fig:x} 
are in reasonable agreement with the preliminary
data of the H1 Collaboration~\cite{h12003,Lemrani:2003mj} (not shown), 
although some differences
are present, especially for $E_T^{\gamma}$ values below 6.7 GeV 
and for negative $\eta_{jet}$.

For $5<E_T^{\gamma}<6.7$ GeV and for $\eta_{jet} < -0.3$
the NLO result are higher than the measured cross section by two standard 
deviations. Smaller differences are also observed
at positive $\eta_{\gamma}$ values, where the predictions
tend to overshot the data. 
In these kinematic ranges the predictions are in good
agreement with the preliminary data, if one takes the number of active 
flavours $N_f=3$ instead of $N_f=4$~\footnote{The previous comparison 
with the ZEUS data~\cite{Breitweg:1999su} for the $ep\ra e\gamma X$
reaction led to opposite conclusions: the predictions for
$N_f=3$ were too low in each kinematic range and better
agreement with data
was observed for $N_f=4$ and $N_f=5$~\cite{Krawczyk:2001tz}.}, 
but the scale $\mu=E_T^{\gamma}$, 
for $E_T^{\gamma}$ above 5 GeV, seems to be too large
to neglect the charm contribution.

Now, we make short comparison with the NLO predictions of Fontannaz, 
Guillet and Heinrich (FGH)~\cite{Fontannaz:2001nq} presented 
in~\cite{h12003,Lemrani:2003mj}.
The FGH calculation takes into account the QCD corrections to the 
resolved-photon processes, which are not included in the 
calculation presented herein (see sec.~\ref{nlo} and 
ref.~\cite{Krawczyk:2001tz}). 
They also use parton parametrizations different than the GRV ones used by us.

In the considered kinematic range
the total cross section of FGH is about 6\% lower than our
predictions. However differences are larger for the differential
cross sections. For example, the FGH cross sections are lower
by about 20\% for $d\sigma /dE_T^{\gamma}$ and $d\sigma /d\eta^{jet}$ 
at $E_T^{\gamma}<6.7$ GeV and $\eta^{jet} < -0.3$,
respectively, and they give better agreement with the data.
On the other hand, our predictions for
$d\sigma /d\eta^{jet}$ ($d\sigma /dx_{\gamma}^{obs}$) 
slightly better describes 
the data at $\eta_{jet} > 1.6$ ($0.25<x_{\gamma}^{obs}<0.5$),
where FGH predictions are about 45\% (13\%) higher than ours.
In other kinematic ranges the differences between
calculations are not larger than $\pm$15\% (usually below 10\%), 
and both calculations
lead to a similar description of the preliminary H1 data, although
the $d\sigma /dE_T^{jet}$ distribution of FGH is a bit 
more flat and more consistent with data in shape.

\section{Summary}\label{sum}

We presented the results of the NLO calculation of the cross section
for the photoproduction of the prompt photon and prompt photon plus jet 
at the $ep$ collider DESY HERA. 

A new exact calculation for the prompt photon production was performed.
We found that the predictions agree within 2\% with the results obtained 
in the small cone approximation used in our previous 
analysis~\cite{Krawczyk:2001tz}.

The main aim of this paper was a detailed analysis of the
prompt photon plus jet production in NLO.
(In the previous analysis~\cite{Krawczyk:2001tz} we presented only the 
LO predictions.) The dependence on the choice of parton distribution 
functions was found to be of order 10\%.
The uncertainties due to the variation of the
renormalization/factorization scale are of order $\pm$ 5\%,
and it may indicate that the corrections of higher orders
are small. But for relatively low $E_T^{\gamma}$,
$E_T^{\gamma} < E_{T,min}^{jet}$, the NLO corrections
are very large while the LO contribution is suppressed,
and we expect that in this region the corrections
of higher orders could change the predictions.

Our predictions are in reasonably agreement with the H1
preliminary data~\cite{h12003,Lemrani:2003mj}, 
nevertheless in some kinematic ranges there are discrepancies
up to two standard deviations.

Differences between our predictions and the predictions
of Fontannaz, Guillet and Heinrich~\cite{Fontannaz:2001nq}
presented in~\cite{h12003,Lemrani:2003mj}
are usually below 10\%, but in some kinematic limits they are larger.
The largest differences, up to 45\% at large $\eta_{jet}$, 
are seen for the $d\sigma /d\eta_{jet}$ distribution.
The experimental errors are too large to conclude which
calculation gives better description of the H1 preliminary data.

~\\

{\bf Acknowledgments}

We would like to thank S. Chekanov from the ZEUS Collaboration
and J. Gayler from the H1 Collaboration for helpful discussions.
We are also grateful to M. Fontannaz for providing fortran
subroutines computing the AFG, AFG02 and BFG parton distributions.
This work was partly supported by the European Community's
Human Potential Programme under contract HPRN-CT-2002-00311 EURIDICE.

\appendix

\newpage
~\\
{\Large\bf Appendix}
\section{The notation}\label{notation}
\begin{figure}[h]
\vskip 6.5cm\relax\noindent\hskip 0cm
       \relax{\includegraphics{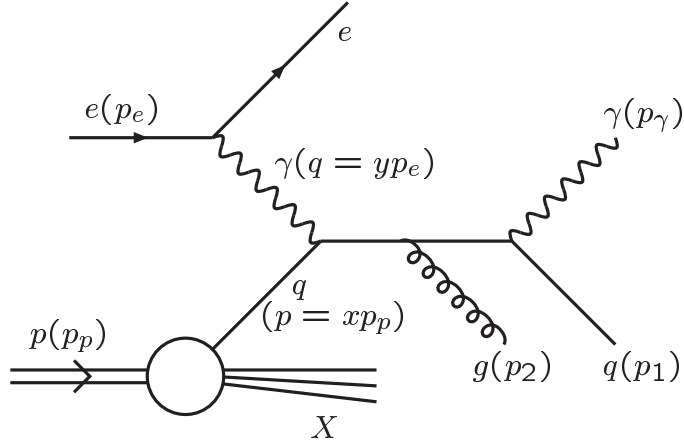}}
\vspace{-0.5cm}
\caption{\small\sl An example of the $2\ra 3$ NLO process. 
The omission of the final gluon gives the lowest order
Born diagram ($\gamma q\ra\gamma q$).}
\label{23}
\end{figure}
Here we introduce the notation which is used in Appendix~\ref{col}
and in sec.~\ref{pss}. There are two kinds of $2\ra 3$ partonic 
processes:
\be
\gamma (q) + q(p) \ra \gamma (p_{\gamma}) + q(p_1) + g(p_2)
\makebox[0.5cm]{} {\rm (see\makebox[0.3cm]{} fig.~\ref{23}),}
\ee
and
\be
\gamma (q) + g(p) \ra \gamma (p_{\gamma}) + q(p_1) + \bar{q}(p_2).
\ee
The four-momenta of particles contributing to the processes
are given in brackets. Let us define variables $v$ and $w$:
\be
v=1+t/s\makebox[1cm]{,}w=-u/(t+s),
\ee
where
\be
s=(q+p)^2\makebox[1cm]{,}t=(q-p_{\gamma})^2\makebox[1.5cm]{and} 
u=(p-p_{\gamma})^2.
\ee
The $v$ and $w$ variables are in the range $0<v<1$ and $0<w<1$.
Note, that for massless particles $(p_1+p_2)^2=sv(1-w)$, and
in the limit $(p_1+p_2)^2\ra 0$ one obtains $w\ra 1$
($v$ can not be too low, since the final photon has large
transverse momentum). For the $2\ra 2$ processes $w=1$ by definition.
 
\section{Cross sections of $2\ra 3$ processes 
for collinear configurations}\label{col}

In this Appendix we have collected analytical formulae for 
the QCD corrections to the Born process
in the regions of the phase space labeled 
as parts 2, 3 and 4 (sec.~\ref{pss}).
The formulae are derived with an assumption that parameter $\theta_{cut}$
($R_{cut}$) is small, $0<\theta_{cut}\ll 1$ ($0<R_{cut}\ll 1$),
and all terms of order ${\mathcal{O}}(\theta_{cut}^n)$ 
(${\mathcal{O}}(R_{cut}^n)$)
are neglected for $n\ge 1$. All the collinear singularities are 
factorized into appropriate parton distributions with use of the
dimensional regularization.

The QCD corrections to the Born process have contributions from all
parts of the phase space:
\be
E_{\gamma}{d\sigma^{\gamma p\ra\gamma (jet) X}
\over d^3p_{\gamma}}_{| QCD~corr}=
\sum_{i=1,2,3,4,5}
E_{\gamma}{d\sigma^{\gamma p\ra\gamma (jet) X}_i
\over d^3p_{\gamma}}_{| QCD~corr}.
\ee
As explained in sec.~\ref{pss}, in parts 1 and 5 we use formulae 
taken from ref.~\cite{Aurenche:1984hc}.
The corresponding cross sections for the $\gamma p\ra\gamma (jet) X$ 
reaction in parts 2, 3 and 4 consists of the contributions
of the $\gamma q\ra\gamma qg$ and
$\gamma g\ra\gamma q\bar{q}$ processes:
\bea
E_{\gamma}{d\sigma^{\gamma p\ra\gamma (jet) X}_{i}
\over d^3p_{\gamma}}_{| QCD~corr.} =
\nonumber 
\eea
\be\label{sumi}
\makebox[0cm]{}
= \int_0^1 \theta(s+t+u)
\sum_{q=u,\bar{u}...}^{2N_f} \left [ f_{q/p}(x)
E_{\gamma}{d\sigma^{\gamma q\ra\gamma qg}_i\over d^3p_{\gamma}}+
f_{g/p}(x)
E_{\gamma}{d\sigma^{\gamma g\ra\gamma q\bar{q}}_i\over d^3p_{\gamma}}
\right ] dx ~, 
\ee
where i= 2, 3 or 4 and the summation runs over all quarks' and antiquarks' 
flavours.
We include $2 N_f$ flavours in $d\sigma^{\gamma g\ra\gamma q\bar{q}}_{2,3,4}$, 
since there are $N_f$ possible pairs $q\bar{q}$, and in each 
pair the quark or antiquark can be collinear to the initial electron, 
initial proton or to the final photon.
The expressions for the partonic cross sections in parts 2, 3 and 4 are:
\be
E_{\gamma}{d\sigma^{\gamma q\ra\gamma qg}_2\over d^3p_{\gamma}}=
{1\over vws^2}
{\alpha\over 2\pi} P_{\gamma\ra q\bar{q}}(w,yE_e\theta_{cut})
|\bar{M}^{q\bar{q}\ra\gamma g}(v)|^2 ~,
\ee
\be
E_{\gamma}{d\sigma^{\gamma g\ra\gamma q\bar{q}}_2\over d^3p_{\gamma}}=
{1\over vws^2}
{\alpha\over 2\pi} P_{\gamma\ra q\bar{q}}(w,yE_e\theta_{cut})
|\bar{M}^{qg\ra\gamma q}(v)|^2 ~,
\ee
\be
E_{\gamma}{d\sigma^{\gamma q\ra\gamma qg}_3\over d^3p_{\gamma}}=
{1\over (1-v)s^2}
{\alpha_S\over 2\pi} P_{q\ra qg}({1-v\over 1-vw},xE_p\theta_{cut})
|\bar{M}^{\gamma q\ra\gamma q}(vw)|^2 ~,
\ee
\be
E_{\gamma}{d\sigma^{\gamma g\ra\gamma q\bar{q}}_3\over d^3p_{\gamma}}=
{1\over (1-v)s^2}
{\alpha_S\over 2\pi} P_{g\ra q\bar{q}}({1-v\over 1-vw},xE_p\theta_{cut})
|\bar{M}^{\gamma q\ra\gamma q}(vw)|^2 ~,
\ee
\newpage
\bea
E_{\gamma}{d\sigma^{\gamma q\ra\gamma qg}_4\over d^3p_{\gamma}}=
\nonumber
\eea
\be
\makebox[0cm]{}
={1\over (1-v+vw)s^2}
{\alpha\over 2\pi} P_{q\ra\gamma q}(1-v+vw,E_T^{\gamma}R_{cut})
|\bar{M}^{\gamma q\ra qg}({vw\over 1-v+vw})|^2 ~,
\ee
\bea
E_{\gamma}{d\sigma^{\gamma g\ra\gamma q\bar{q}}_4\over d^3p_{\gamma}}=
\nonumber
\eea
\be
\makebox[0cm]{}
={1\over (1-v+vw)s^2}
{\alpha\over 2\pi} P_{q\ra\gamma q}(1-v+vw,E_T^{\gamma}R_{cut})
|\bar{M}^{\gamma g\ra q\bar{q}}({vw\over 1-v+vw})|^2 ~,
\ee
where we have used the notation:
\be
P_{g\ra q\bar{q}}(z,E) = {1\over 2} \left\{
[z^2+(1-z)^2] \ln{(1-z)^2E^2\over \mu^2} +1 \right\} ~,
\ee
\be
P_{\gamma\ra q\bar{q}}(z,E) = 2 N_C e_q^2 P_{g\ra q\bar{q}}(z,E) ~,
\ee
\be
P_{q\ra qg}(z,E) = C_F \left\{
{1+z^2\over 1-z}\ln{(1-z)^2E^2\over \mu^2} +1-z \right\} ~,
\ee
\be
P_{q\ra\gamma q}(z,E) = e_q^2 \left\{
{1+(1-z)^2\over z}\ln{(1-z)^2E^2\over \mu^2} +z \right\} ~,
\ee

\be
|\bar{M}^{\gamma g\ra q\bar{q}}(v)|^2 = 
\alpha\alpha_S e_q^2 {v^2+(1-v)^2\over v(1-v)} ~,
\ee
\be
|\bar{M}^{q\bar{q}\ra \gamma g}(v)|^2 =
{2C_F\over N_C} |\bar{M}^{\gamma g\ra q\bar{q}}(v)|^2 ~,
\ee
\be
|\bar{M}^{\gamma q\ra\gamma q}(v)|^2 =
2\alpha^2e_q^4{1+v^2\over v} ~,
\ee
\be
|\bar{M}^{\gamma q\ra qg}(v)|^2 =
{C_F\over e_q^2}{\alpha_S\over\alpha}
|\bar{M}^{\gamma q\ra\gamma q}(1-v)|^2 ~,
\ee
\be
|\bar{M}^{qg\ra\gamma q}(v)|^2 =
{1\over 2 N_C e_q^2}{\alpha_S\over\alpha}
|\bar{M}^{\gamma q\ra\gamma q}(1-v)|^2 ~.
\ee



~\newpage
\vspace*{7.cm}
\begin{figure}[ht]
\vskip 17cm\relax\noindent\hskip -2cm
       \relax{\includegraphics{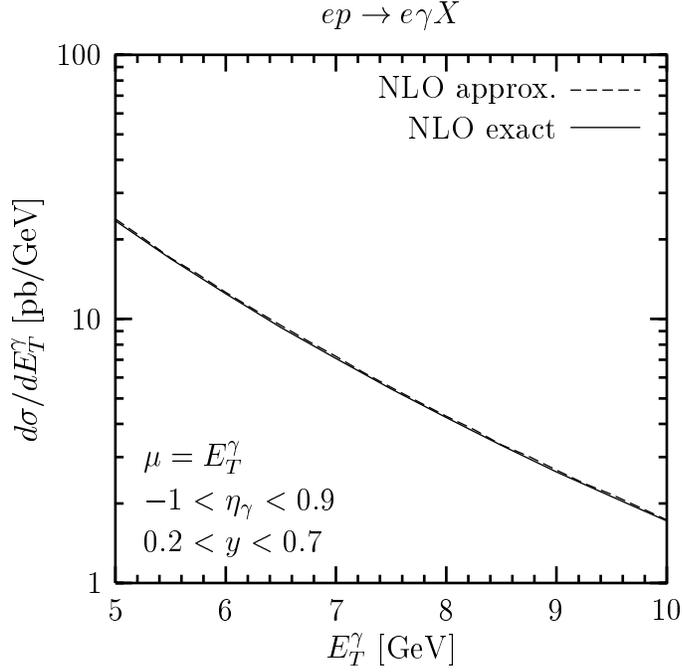}}
\vspace{-15.5cm}
\vskip 24.5cm\relax\noindent\hskip -2cm
       \relax{\includegraphics{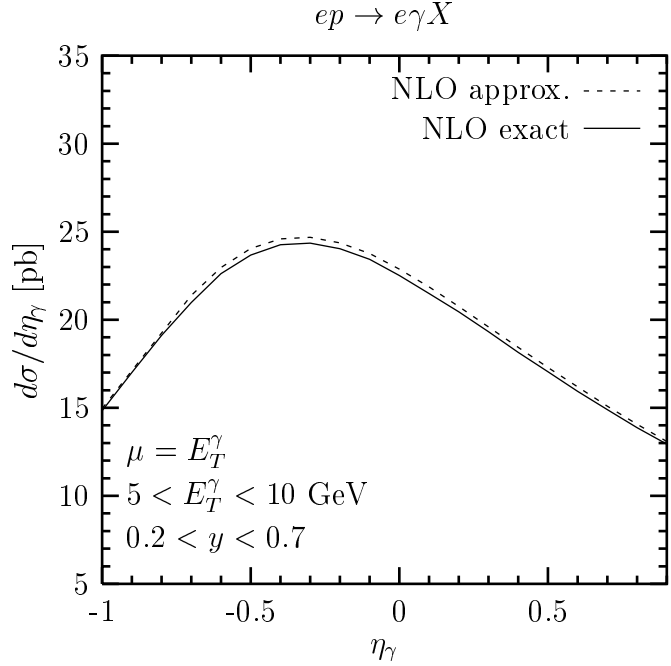}}
\vspace{-16.cm}
\caption{\small\sl The comparison between predictions 
obtained in the small cone approximation (dashed line) and exact 
ones (solid line)
for the $ep\ra e\gamma X$ photoproduction.
The $d\sigma /dE_T^{\gamma}$ (a) and
$d\sigma /d\eta_{\gamma}$ (b) cross sections are shown.}
\label{fig:nojet}
\end{figure}

\vspace*{7.cm}
\begin{figure}[ht]
\vskip 17cm\relax\noindent\hskip -2cm
       \relax{\includegraphics{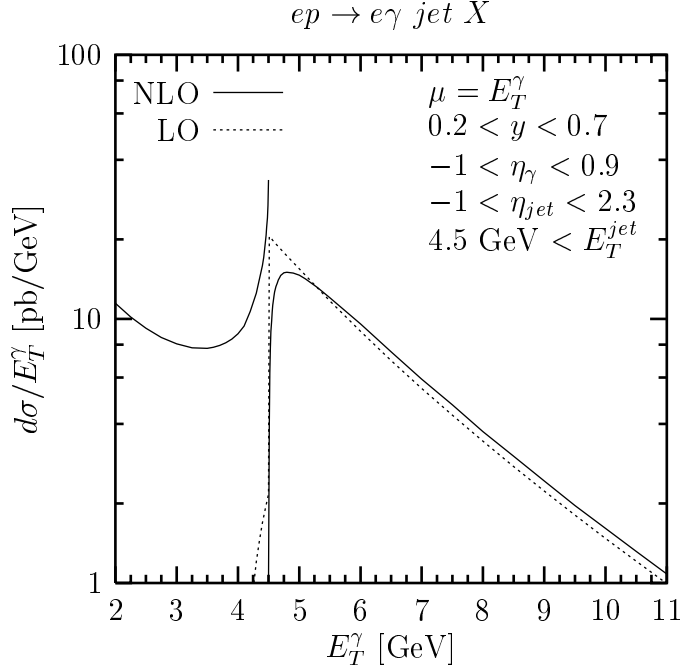}}
\vspace{-15.5cm}
\vskip 24.5cm\relax\noindent\hskip -2cm
       \relax{\includegraphics{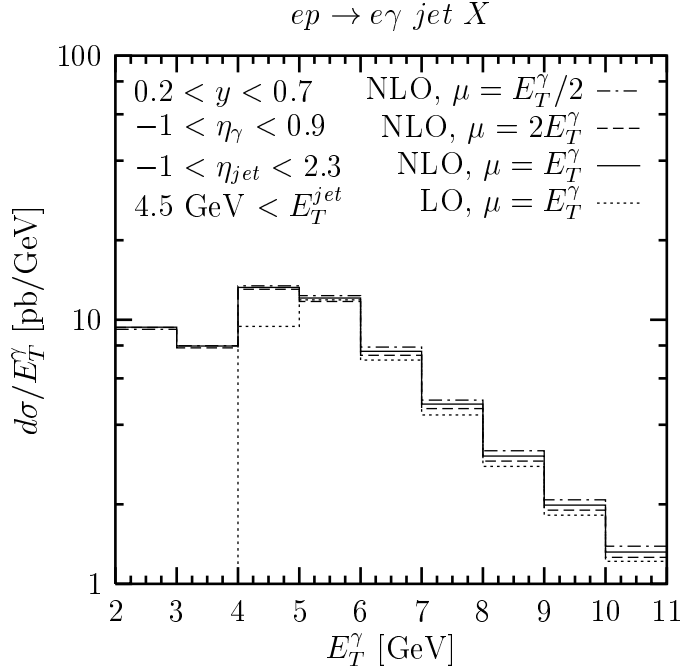}}
\vspace{-16.cm}
\caption{\small\sl The differential cross section $d\sigma /dE_T^{\gamma}$ 
(a) and the differential cross section $d\sigma /dE_T^{\gamma}$ averaged
over $E_T^{\gamma}$-bins (b) for the $ep\ra e\gamma ~jet ~X$ process.
The NLO predictions for $E_T^{\gamma}=\mu$ (a) and for $\mu$ between 
$E_T^{\gamma}/2$ and $2E_T^{\gamma}$ (b) are shown together with 
the LO predictions.}
\label{fig:ptgamma}
\end{figure}

\vspace*{7.cm}
\begin{figure}[ht]
\vskip 17cm\relax\noindent\hskip -2cm
       \relax{\includegraphics{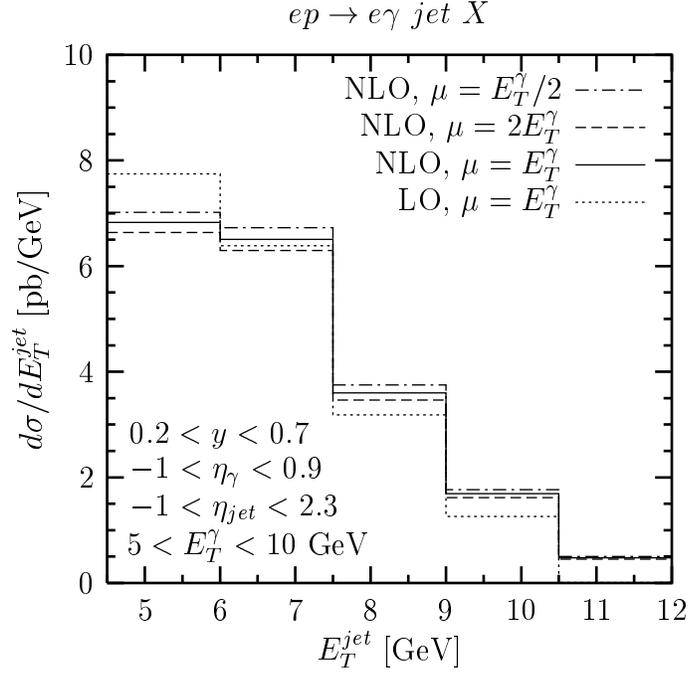}}
\vspace{-15.5cm}
\vskip 24.5cm\relax\noindent\hskip -2cm
       \relax{\includegraphics{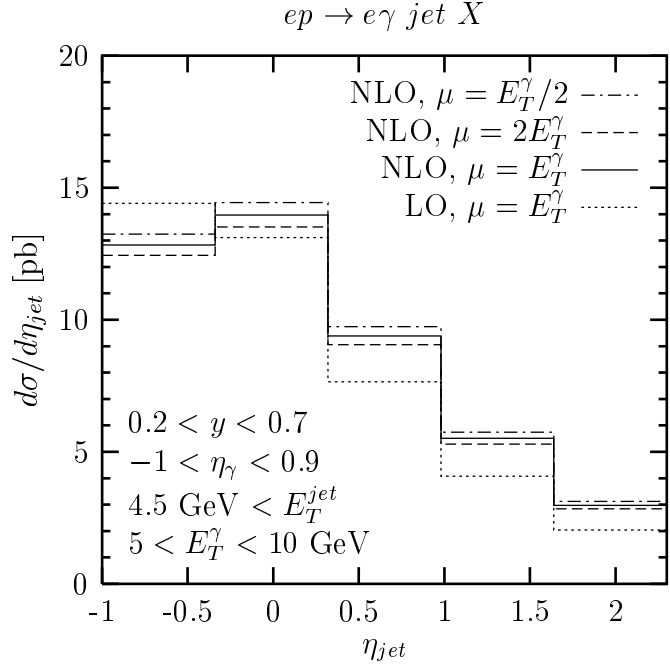}}
\vspace{-16.cm}
\caption{\small\sl 
The NLO predictions with $\mu$ between $E_T^{\gamma}/2$ and $2E_T^{\gamma}$
and the LO predictions for $d\sigma /dE_T^{jet}$ (a) and $d\sigma /d\eta^{jet}$
(b) distributions.}
\label{fig:jet}
\end{figure}

\vspace*{7.5cm}
\begin{figure}[ht]
\vskip 17.cm\relax\noindent\hskip -2cm
       \relax{\includegraphics{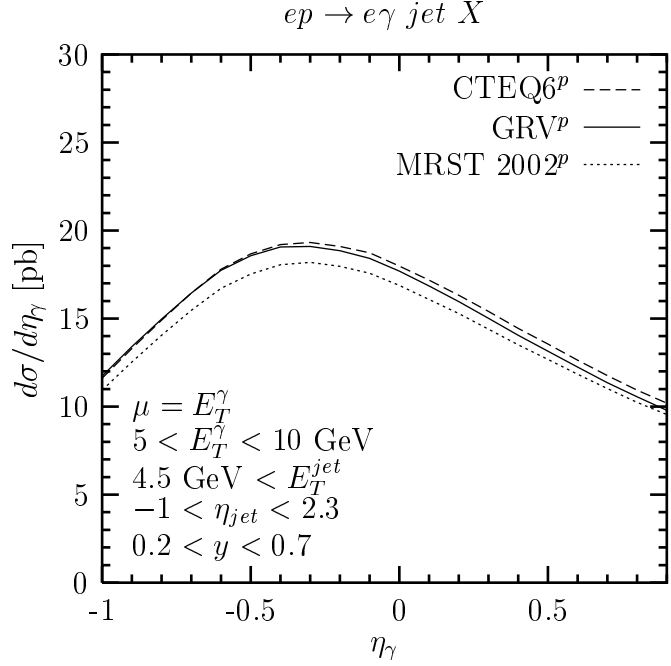}}
\vspace{-16cm}
\caption{\small\sl The differential cross section $d\sigma /d\eta^{\gamma}$
for GRV~\cite{Gluck:1995uf}, MRST2002~\cite{Martin:2002aw} and
CTEQ6~\cite{Pumplin:2002vw} parton distributions in the proton.}
\label{fig:etagamma}
\end{figure}

\vspace*{7.5cm}
\begin{figure}[ht]
\vskip 17.cm\relax\noindent\hskip -2cm
       \relax{\includegraphics{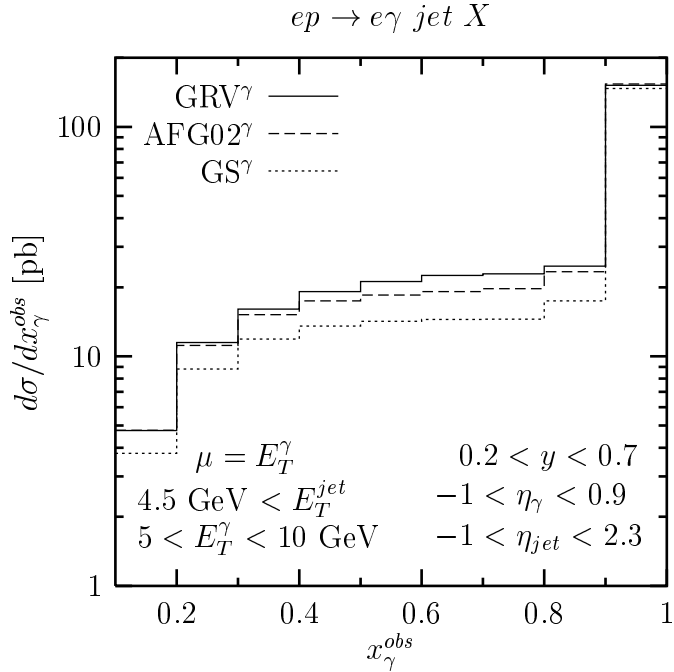}}
\vspace{-16cm}
\caption{\small\sl The differential cross section $d\sigma /dx_{\gamma}^{obs}$
for GRV~\cite{Gluck:1992ee}, GS~\cite{Gordon:1997pm}
and AFG02~\cite{afg02} parton distributions in the photon.}
\label{fig:x}
\end{figure}

\end{document}